\documentstyle[preprint,aps]{revtex}
\begin{document}
\draft
\title
{Distributions of Triplets in Genetic Sequences }
\author{Yu Shi\footnote{corresponding author, email: shiyu@alon.cc.biu.ac.il},
Ido Kanter and David Kessler}
\address{Minerva Center and
Department of Physics, Bar-Ilan University,
Ramat Gan 52900, Israel
}
\maketitle
\begin{abstract}
Distributions of triplets  in some genetic sequences are examined and
found to be well
described by a 2-parameter  Markov process with  a sparse
transition matrix. 
The variances of all the relevant parameters are
 not large, indicating that most sequences gather in a small region in
 the parameter space. Different sequences have very similar  values of 
  the entropy calculated directly from the data and 
   the two parameters characterizing the 
 Markov process fitting the sequence.
 No  relevance  with taxonomy or coding/noncoding
 is clearly observed. 
  
 \vspace{1cm}
Keywords: Markov process, genetic sequence, DNA
\end{abstract}

\pacs{PACS numbers: 87.10.+e, 02.50.Ga, 05.40.+j}

\section{INTRODUCTION}
 In recent years, 
 methods of statistical mechanics are applied in
 other fields of research 
 based on mapping the quantities 
 under study to physical or numerical quantities, e.g.
 spins or binary numbers 
``0'' and ``1'',
 from which
 various measures can be calculated and analysed \cite{china,peng}.
  Such is the case in
  the recent investigations on   
the statistical properties
 of   DNA sequences    and  human languages
 \cite{peng,li,schenkel,mantegna,shtrikman,ebeling,kanter},
 as well as music \cite{boon,shi}.
  The observation that
  local grammar-like rules affect the global statistical nature of
sequences is in accordance  with the philosophy of statistical mechanics.

An interesting  issue concerns the   distribution 
of semantic units ``words''; words in a languange 
 and the 64 triplets (3-tuples) 
in genetic sequences.  
The frequency of the occurence of each semantic
 unit is calculated, and  the units are ordered in a
  the decreasing order
 of frequency, $P(1)\,\geq\, P(2)\,\geq\,\cdots\,\geq\,P(N)$, where $N$ is 
 the size of the vocabulary.
 For languages, there was a so-called
  Zipf's law that 
  \begin{equation}
  P(k)\,\propto\,k^{-\rho}
  \end{equation}
  with $\rho\,\sim\,1.0$ \cite{zipf}. 
   In
  DNA sequences, triplets in coding regions are the ``words'',
  since the coding regions are transcribed to RNA, where the 
  nonoverlapping triplets code the amino acids.
  It is unknown whether there are ``words'' in noncoding regions.
  Recently, distributions of n-tuples ($n$ ranges from $3$ to $8$)
  were analysed and it was claimed that Zipf's law holds and  that
 $\rho$ is consistently larger for the noncoding
  sequences than coding sequences and therefore
   the former are more similar 
  to languages \cite{mantegna}. This conclusion
  was  heavily 
  criticized \cite{israeloff}.
In fact, though it was 
    appealing due to the earlier attempts to relate it to the 
 structure of language \cite{mandelbrot}, Zipf's law
 had
  been  acknowledged as ``linguistic shallowness'' since it can be
  generated from random text \cite{man2}\cite{miller}\cite{israeloff},
  it was claimed recently that 
 an initial inverse power law  in the distribution can be obtained
  under quite general 
  conditions \cite{perline}. 		        
     On the other hand, 
     it was pointed out that 
  $\rho\,=\,-d\ln (P)/d\ln(k)$ is,
 in fact, a increasing function of $k$,  there is no macroscopic 
 regime  where $\rho$ is a constant, consequently 
 any attempt to fit the data with a single $\rho$ is sensitive to the
  details of
the  fitting \cite{kanter}. 

For  the occurrence of letters over the alphabet in  biological sequences as well
as in over 100 human languages,
it was claimed that the ordering of frequencies  
approximates \cite{shtrikman}
\begin{equation}
 P(k)\,=\,A-D\ln (k),	 \label{log}
 \end{equation}
 where  $A$ and $D$ are constants, the normalization condition
 reduces the independent parameter to only one. 
 An exception was found to be
 Chinese, where
  the corresponding distribution is nearer to Zipf's law. This can 
  be understood;
   there is no letter in Chinese unless it is transformed to the
  alphabetic system according to the pronounciation, 
   while the character, which had been  considered  to be the letter
  since it is the basic unit,
   also embeds meanings.  
      	
 The characterization and explanation 
 of the distributions demands a model beyond the Zipf's law. 
 A
2-parameter random Markov Process (MP) was proposed for the generation of these 
 sequences \cite{kanter}, with the observations mentioned above 
 being 
 natural consequences.
  Can the distributions of 3-tuples in various different 
  genetic sequences
  be well described by the MP model?  The  
  positive answer is given by showing that the distributions for
  each sequence that is long enough
  can be fitted very well by a MP with certain parameters, while
 the features for  short sequence are consequences of finite
  Markov chain.  But  no relevance with taxonomy or coding/nocoding issue
  is clearly observed. 
 
The MP model is explained in Sec. II; the analysis on 
genetic 
sequences is reported in Sec. III; Sec. IV contains
 the  conclusions. 
 
\section{THE TWO-PARAMETER MARKOV PROCESS}
 A  Markov process 
 is the simplest algorithm for the stochastic production
 of sequences. Consider the generation of a sequence composed of ``words'' 
 or  states
 chosen from $N$ possibilities. If the probability distribution for choosing
 the next ``word'' is only a function of the current last one, then
 this process can be considered as a MP. The transition probability 
from state $i$ to $j$ is denoted as $W(i,j)$. There is
 the normalization condition $\sum_{j}W(i,j)\,=\,1$.
 The probability of occurrence of each ``word'' in the sequence which
 is long enough is the stationary solution of MP.
 The ingredients of this model are as follows: 
 
(a) Number of states. For simplicity, $N$ is fixed to be equal to
$2^{L}$ with $L$ an integer, each state is identified by an
   $L$-bit binary
  number between $0$ and $N-1$. For the genetic sequences,
  $N\,=\,64$ and thus $L\,=\,6$.
  
(b) Sparseness of $W$. Reflecting the grammatical 
rule, the transition matrix $W(i,j)$ is assumed 
to be sparse, i.e. the
  number of nonvanishing elements in each row, $C$,
  is finite and does not scale with $N$. The simplest nontrivial 
  case is
  $C\,=\,2$. 
  
 (c) Permissible connectivity.
 The cornerstone of the discussed MP is that the transition matrix differs
 from a random graph. In a language,
 for instance, the semantic and grammatical rules require  that
 a word is not
 haphazardly followed by a random selection of other words. Rather,
 the choice of the  successive word is strongly constrained.
 This fact is modeled in the following manner. The two states $m_{0}$
 and $m_{1}$ connected to the state $m$ are given by
\begin{equation}
m_{0}\,=\,(2m) mod (N);\,m_{1}= (2m) mod (N)+1,
\end{equation}
where 
 $m\,=\,0,1,...,2^{L}-1$. In words, the $L-1$ rightmost
 bits of state $m$ are shifted one bit to the
  left, and the rightmost bit is set equal to either $0$ or $1$. 
  Thus each successive word is closely related to the one before,
   the outword and inword connectivity of each state is equal to $2$.
   
   (d) Strength of transition probabilities.
  The two weights, transition probabilities, going from each state take
  the value $x$ and $1-x$.
  
  (e) Bias. Another
  parameter, the bias, is introduced to distinguish the two options. 
  We pick $W(m,m_{0})\,=\,1-x$ and  $W(m,m_{1})\,=\,x$ with probability
  $B$, and vice versa with probability $1-B$.  
  
  The bias can be thought to be
  related to some global constraints by the  ``meaning'' of the text in addition
  to those reflected by the local rules. 
  When  $B\,=\,0.5$, i.e. there is 
   no bias, this MP was found \cite{kanter} to lead to 
    a  distribution  approximating log-normal rule, Eq. (\ref{log}), which 
 is held quite well by the distribution of letters. This
 can be understood through  the fact that  
  the sequence of letters is only restricted by local phonetic preferences.
  									        
  Because of  global
  inversion symmetries $x\,\rightarrow\,1-x$ and
   $B\,\rightarrow\,1-B$, the interesting regime in the unit square of
   $(x,B)$ may only be $(0.5 - 1, 0.5 - 1)$.  Furthermore,
    changing only
   $x$ to $1-x$, or $B$ to $1-B$
   is only changing the role between $0$ and $1$.

 An important variable is   the
   average drift towards 1,
   $x_{eff}\,=\, xB\,+\, (1-x)(1-B)$. 
It was found that a function obtained 
   by rescaling the local
   slope of the distribution function only depends on $x_{eff}$ in addition
   to the rank order \cite{kanter}.   
    For $x\,=\,0.5$, we have    $x_{eff}\,=\,0.5$ independent of
  $B$. Another interesting quantity is the  Markov entropy 
 \begin{equation}
S_{m}\,=\,-x\ln x-(1-x)\ln(1-x),
\end{equation}
which is independent of $B$.
 
 The feature of this type of MP model have
  been found to be robust for many modifications. For example,
 the qualitative feature does not change if dependence of the next 
  state on more former states than the current last is introduced, or
  higher but still finite connectivities are allowed. It was 
  also found
  that all the distributions resulting from these extended models could be
  readily mapped to the simplest one \cite{halibard}.
Therefore this two-parameter model can serve as a prototypical
model  even for 
less sparse matrix, which might possess many parameters.
  
\section{GENETIC SEQUENCES}
Genetic sequences of different taxonomic divisions
 are randomly selected from GenBank Release
No. 97 \cite{gene}. First,  for short sequences there are, of course, 
many plateaus in the ordered distribution of triples,
 and cannot be 
fitted by the stationary solutions of the  MP process. 
This is a finite-size effect and just a support for the validity
 of this model. 
 As an example, compare  the
 ordered distributions for  bacteriophage P1 gene10 with
 1127 bp as shown  in Figure \ref{finite} (a) with
  the ordered 
 distribution generated  by a MP with $x\,=\,0.69$, $B\,=\,0.62$
 after $500$ steps as shown in Figure \ref{finite} (b).
 
 We analysed in detail
22 long sequences: the longest one being 
 s. cerevisiae chrosome III complete DNA sequence with 315341 bp;
    the shortest one comprises 6061 bp;
     6 sequences are complete DNA genome-s;
      5 are complete cds-s, i.e.,
       sequences coding for amino acids in protein;
  3 RNA sequences, 
 2 of them are complete genome.  Different sequences are listed
 and numbered in Table \ref{tab1}.
 
We  fit the data of the distributions in terms of that 
generated by the MP model.
For each sequence, the distribution of triplets
 is calculated and ordered
in decreasing order. Then the parameters $x$ and $B$ are found
for the best fitting MP with the least value of the
 cost function defined as
\begin{equation}
Cost\,= \, \sqrt{\frac{1}{64}\sum_{k=1}^{64}D^{2}(k)},
\end{equation}
where 
\begin{equation}
D(k)\,=\,\frac{P_{s}(k)-P_{m}(k)}
{P_{s}(i)},
\end{equation}
$P_{s}(k)$ is the rank-ordered distribution of triplets for a
genetic sequence,
$P_{m}(k)$  is the rank-ordered distribution of $6-$bit binary numbers 
for a MP.
In the two 
dimensional lattice parameter  space $(x,B)\,=\,(0.5-1,0.5-1)$
with lattice constant $0.01$, we search
for  the MP which  fits each sequence with
minimal cost.
   Three examples of the 
  distribution and its fitting to MP are shown
  in Figure \ref{fig2}. It can be seen  that  the fitting is quite good.
Such is the case 
   for 16 sequences. For the remaining 6, there
   is a discontinuous decrease at  a high rank $k\,=\,54$ or $56$.
This discontinuity at  the tail 
might be due to fluctuations and does not affect
our general discussions. 
A satisfactory
   fitting can be found eliminating the last several  
   points.
See Figure \ref{fig3} for an illustration. 

Note that our 
 fitting is global instead of being  part of the data,
i.e.,  the contributions to the cost function 
do not come mainly  from the tail, as shown in Figure \ref{fig4}.
 
The quantitative 
results are summarized  in TABLE 2 for all the sequences
we analyzed. We  present the values of the cost,
 $x$
and $B$. From $x$ and $B$ we calculate 
$x_{eff}$ and the Markov entropy $S_{m}$ 
of the corresponding Markov process, and the Shannon 
entropy 
\begin{equation}
S\,=\,-\sum_{k}P_{s}(k)\ln P_{s}(k)
\end{equation}
  calculated directly from the original data of each sequence.
  For a completely random sequence,
  $P(k)\,=\,1/64$, thus $S\,=\,\ln(64)\,\approx\,4.1589$.

  It is clear that the costs are very small;
    the largest one is $0.0807$ while 
the least is $0.0273$.
The average  and variance of the results over
 {\em all} sequences are presented in 
Table \ref{tab3}, in addition, the average of  costs is $0.0555$. 
It is remarkable that the  relative variance,
i.e. variance divided by
the average for each quantity, is not large.
In particular, that for $x_{eff}$ is only
$0.0485$, implicating  that $x_{eff}$ is
a very special
 quantity,  while that for $S$, which is 
model-irrelevant, is only $0.0179$. 
The relative variances for $x$, $B$ and 
$S_{m}$ are either not large, though larger 
than those for $S$ and $x_{eff}$.
   It can be seen that the statistics  
are different  but not  far from each other, and that 
most sequences ocuppy a small region in the
parameter space, which is distinct but not very far
 from the complete
randomness with $x\,=\,x_{eff}\,=\,0.5$, $S_{m}\,=\,0.6931$
and $S\,=\,4.1589$.

A problem is whether there is a distinction in quantities
discussed here between coding and noncoding sequences. 
To examine this possibility, we calculate the average values over
sequences No. $7,\,8,\,9.\,10,\,12,\,13,\,19,\,20$. 
These sequences are  complete coding sequences or RNA. Both are
 $100\%$ coding. Comparing  Table \ref{tab4} with Table \ref{tab3},
it can be seen that $x$ is  larger than the average over all sequences,
and $S_{m}$ and $S$ are smaller, clearly in contrast 
to the claim that noncoding regions are more similar to languages than
coding regions \cite{peng}. But the difference is so small that
no definite conclusion can be drawn.
On the other hand,  the differences with those of the language  are
still very large, since values of
$x$ and $B$ were found both to be  $0.92$ \cite{kanter},
 a very large value.
Similar investigations are made on whether 
there is relevance
 between the quantities 
characterizing  sequences and  the different 
taxonomic divisons.
We calculate  the averages and 
variances for 
each division, as listed  in Table \ref{tab5}.
It can be seen that there is no monotonic trend with the evolution.
To examine whether sequences in the same division are 
closer to each other compared with all sequences,
we compare the overall variances in Table \ref{tab3}
and the 
variances for viral and primate in Table \ref{tab5},
since  for other divisions only one or two
sequences  are analysed. It can be seen that
some are
  larger while some are smaller than those for 
all the sequences. Therefore, 
in our result there is
no sign of  relevance between these quantities and taxonomy. 

The  distribution of 
triplets remains nearly unchanged if the starting nucleotide shifts 
$1$ or $2$ behind. This can be seen from  Figure \ref{shift}
 showing the
 distributions for
the original  s. cerevisiae chrosome III complete DNA sequence,
 and those  shifted $1$ and $2$ behind.  This result holds for 
 all sequences.
   
\section{Conclusions}

(1) Statistics of examined
genetic sequences are  well described by 
the 2-parameter Markov process.

(2) Most sequences gather in a small region in the parameter
$(x,B)$ space. The entropy $S$  of the data and 
$x_{eff}$ measured in the MP model 
are very near to each other for different sequences. 

(3) No relevance of the quantities studied here with coding/noncoding issue 
or with taxonomy 
 is observed. 

(4) The distribution of triplets remains unchanged if the sequence
is shifted.
   
More biologically relevant information might
 be exposed  when
the distribution and transition matrix
are analysed according to the real triplets instead of
to the rank order.  In this way,
the transition matrix  varies 
from sequence to sequence, determined by the different biochemical 
enviornments.

\acknowledgements
      Y.S. thanks  BIU for hospitality.
 I.K. an D.A.K
 acknowledge the support of the Israel Academy of Science.

\begin{figure}
\caption{(a) Rank-ordered   distributions of triplets for
 bacteriophage P1 gene10 with
 1127 bp. (b) Rank-ordered   distributions resulted from the
 $2$-parameter Markov process with $x\,=\,0.69$, $B\,=\,0.62$
 after $500$ steps.}
 \label{finite}
% 1a.ps, 1b.ps
 \end{figure}
 
 \begin{figure}
\caption{Rank-ordered distributions of triplets for genetic
sequences and  of the $6$-bit binary numbers for
the $2$-parameter Markov process which best fit  the
sequences. (a) No. 1, (b) No. 15, (c) No. 17. }
\label{fig2}
% 2a.ps 2b.ps 2c.ps
\end{figure}

\begin{figure}
\caption{There is a discontinuity at $k\,=\,54$ in
the distribution of triplets for sequence No. 10. A 
$2-$parameter Markov process can be found to give  a
satisfactory fit if the last $10$ points are 
neglected.}
\label{fig3}
%3.ps
\end{figure}

\begin{figure}
\caption{The relative difference between the rank-ordered distribution
of triplets in sequence No. 1 (Bacteriophage lambda)
and that of $6$-bit binary numbers in the
$2$-parameter Markov process giving the best fit $D(k) = 
[P_{s}(k)-P_{m}(k)]/P_{s}(i)$.}
\label{fig4}
%4.ps
\end{figure}

\begin{figure}
\caption{Rank-ordered distributions of triplets for 
 (a) original s. cerevisiae chrosome III complete DNA sequence,
(b) shifted $1$ behind, (c) shifted $1$ behind. They are very near
to each other.}
\label{shift}
\end{figure}
 
\begin{table}
\caption{Information on the $22$ sequences analysed in this
paper, they are numbered for the convenience 
of presenting the results. No. 7, 8, 9 are RNA; all others
are DNA.
\label{tab1}}
\begin{tabular}{lllcr}
 No.  &    Locus name   &  Definition   & Taxonomic division   &  Length (bp) \\
\hline
1    &	   LAMCG   & Bacteriophage lambda & phage &  48502 \\ 
     &             &   complete genome	   &	   &	    \\
2& MYP4CG  &Bacteriophage P4 &phage & 11624 \\
 &&complete  genome&&\\
3&HSECOMGEN &Equine herpesvirus, &viral&150223 \\
 &&complete genome&&\\
4&VACRHF&Vaccinia virus genomic DNA&viral& 42090 \\
5&ASFV55KB&  African swine fever virus  &viral& 55098 \\
6& HEHCMVCG &Human Cytomegalovirus Strain AD169 &viral& 229354\\
   &             &   complete genome	   &	   &	    \\
7&TOEAV & Equine arteritis virus (EAV)& viral&12687 \\
&             & RNA genome  &	   &	    \\
8& WNFCG   & West Nile virus RNA & viral&10960 \\
    &             &   complete genome (RNA)	   &	   &	    \\
9&FIVPPR & Feline immunodeficiency virus&viral&9468 \\
   &             &   complete genome (RNA)	   &	   &	    \\
10& RTUORFS   & Rice tungro bacilliform virus &viral&8000 \\
    &             &   complete cds	   &	   &	    \\
11& CSHCG   &Cacao swollen shoot virus polyprotein gene	 &viral& 7161 \\
  &             &   complete circular genome	   &	   &	    \\
12& SBVORFS & Sugarcane bacilliform virus&viral&  7568 \\
  &              &   complete cds	   &	   &	    \\
13  & ANAAZNIF  &Anabaena azollae nifB operon&bacterial&  6061 \\
  &              &   complete cds	   &	   &	    \\
14&SCCHRIII   &   S.cerevisiae chromosome III &plant& 315341 \\
   &              &   complete DNA sequence	   &	   &	    \\
15&  TGDNAPRRA &T.godoii (strain P)&invertebrate&  8350 \\
16 & TGDNARH &T.gondii (RH) &invertebrate&8352 \\
17& MMCOL3A1  & M.musculus COL3A1 gene for collagen alpha-I & rodent& 43601 \\
18& PTMITG  & P.troglodytes mitochondrial DNA & primate& 16561 \\
   &             &   complete genome (isolate Jenny)	   &	   &	    \\
19& HUMCFVII  &  Human blood coagulation factor VII gene  & primate&  12850 \\
   &              &   complete cds	   &	   &	    \\
20& HUMRETBLAS &   Human retinoblastoma susceptibility gene & primate& 180388 \\
  &              &   complete cds	   &	   &	    \\	  
21& HUMHBB &  Human beta globin region on chromosome 11 & primate&  73308\\
22& HSP53G &  Human p53 gene  & primate&   20303 \\ 			
  \end{tabular}
\end{table}

\begin{table}
\caption{Quantitative results on the $22$ sequences.
 $S$ is the entropy of 
the sequences, $cost$ is  a measure of the fitting,
$x$ and $B$ characterize the Markov process giving
least $cost$, $x_{eff}$ is a function of $x$ and $B$,
the Markov entropy $S_{m}$  is a function of $x$. 
See the text for  definitions.
}
\begin{tabular}{lcccccc}
No.&$cost$&$x$&$B$&$x_{eff}$&$S_{m}$&$S$\\
\tableline
1&0.0689&0.6100&0.6800&0.5396&0.6687&4.1225\\
2&0.0491&0.6900&0.5100&0.5038&0.6191&4.0891\\
3&0.0273&0.6000&0.7000&0.5400&0.6730&4.1201\\
4&0.0654&0.7600&0.6100&0.5572&0.5511&3.9591\\
5&0.0807&0.6200&0.8300&0.5792&0.6640&4.0475\\
6&0.0441&0.6100&0.7700&0.5594&0.6687&4.1073\\
7&0.0737&0.7000&0.5400&0.5160&0.6109&4.0949\\
8\tablenotemark[1]&0.0355&0.7500&0.5900&0.5450&0.5623&4.0005\\
9\tablenotemark[1]&0.0643&0.7100&0.7400&0.6008&0.6022&3.9256\\
10\tablenotemark[1]&0.0726&0.8100&0.5600&0.5372&0.4862&3.8553\\
11&0.0646&0.7400&0.5200&0.5096&0.5731&4.0356\\
12&0.0543&0.7400&0.6500&0.5720&0.5731&3.9616\\
13&0.0484&0.6800&0.6200&0.5432&0.6269&4.0545\\
14&0.0404&0.7000&0.5200&0.5080&0.6109&4.0629\\
15&0.0322&0.6000&0.7400&0.5480&0.6730&4.1131\\
16&0.0309&0.6100&0.6800&0.5396&0.6687&4.1139\\
17\tablenotemark[1]&0.0630&0.7200&0.5100&0.5044&0.5930&3.9761\\
18&0.0716&0.7900&0.5300&0.5173&0.5140&3.9838\\
19&0.0673&0.7600&0.6100&0.5572&0.5511&3.9829\\
20\tablenotemark[1]&0.0590&0.7000&0.5300&0.5120&0.6109&3.9991\\
21\tablenotemark[1]&0.0638&0.6600&0.5700&0.5224&0.6410&4.0181\\
22&0.0709&0.6700&0.6700&0.5578&0.6342&4.0774\\
\end{tabular}
\tablenotetext[1]{There is a discontinuity at rank order
$k\,=\,54$ in the rank-ordered distribution of triplets for sequence
No. 10, and at $k=56$ for sequences No. 8, 9, 17, 20, 21.
A Markov process fitting  each of them
satisfactorily can be found 
if the points after the discontinuity are neglected.} 
\label{tab2}\end{table}

\begin{table}
\caption{The average value, variance and relative 
variance of the five quantities
calculated over all the $22$ sequences analysed.
\label{tab3}}
\begin{tabular}{cccc}
quantity &average&variance&variance/average\\
\tableline
     $x$& 0.6923& 0.0640& 0.0924\\
     $B$& 0.6218& 0.0942& 0.1515\\
$x_{eff}$&0.5395& 0.0261& 0.0485 \\
$S_{m}$&  0.6080& 0.0534& 0.0878\\
     $S$& 4.0319& 0.0720 &0.0179\\
\end{tabular}
\end{table}

\begin{table}
\caption{The average value, variance and relative 
variance of the five quantities
calculated over the RNA sequences No. 7, 8, 9,
and the complete coding DNA sequences No. 10,
12, 13, 19, 20.}
\begin{tabular}{cccc}
quantity &average&variance&variance/average\\
\tableline
$x$& 0.7312& 0.0422& 0.0578\\
$B$& 0.6050& 0.0682& 0.1128\\
$x_{eff}$& 0.5479& 0.0291&0.0531 \\
$S_{m}$&  0.5779& 0.0456& 0.0789\\
$S$& 3.984& 0.0739 &0.0186\\
\end{tabular}
\label{tab4}
\end{table}

\begin{table}
\caption{The average value of the five quantities
calculated over each taxonomic division, the variances and 
the relative variances are also given for 
the divisions with more than one
 sequence analysed here.
 The number within the parentheses after each division name
 is that of the analysed sequences belonging to this division.
 \label{tab5}}
\begin{tabular}{lcccc}
division&quantity &average&variance&variance/average\\
\tableline
phage (2)   &$x$&      0.6500&0.0566& 0.0870\\
        &$B$&      0.5950& 0.1202& 0.2020\\
	&$x_{eff}$&0.5217& 0.0253& 0.0485\\
	&$S_{m}$& 0.6439& 0.0351& 0.0545\\
	&$S$&     4.1058& 0.0236& 0.0057\\
viral (10) &$x$& 0.7040& 0.0714& 0.1014\\
       &$B$&0.6510& 0.1052& 0.1617\\
       &$x_{eff}$&0.5516& 0.0281& 0.0509\\
      	&$S_{m}$& 0.5965& 0.0600& 0.1005\\
	 &$S$&   4.0107& 0.0862& 0.0215\\
bacteria (1) &$x$&   0.6800&       &   \\ 
  &$B$&0.6200&&\\
   &$x_{eff}$& 0.5432 &&\\
   	&$S_{m}$&0.6269&&\\
      &$S$& 	4.0545&&\\
plant (1) &$x$&       0.7000&&\\
  &$B$&0.5200&&\\
  &$x_{eff}$& 0.5080&&\\
  &$S_{m}$& 0.6109&&\\
  &$S$&    4.0629&&\\
 invertebrate (2) &$x$& 0.6050& 0.0071 &0.0117\\
   &$B$& 0.7100& 0.0424& 0.0598\\
  &$x_{eff}$&   0.5438&0.0059& 0.0109\\
 &$S_{m}$&  0.6709& 0.0030& 0.0045\\
 &$S$&  4.1135& 0.0005& 0.0001\\
 rodent (1) &$x$&  0.7200&&\\
&$B$&  0.5100&&\\
&$x_{eff}$&0.5044&&\\
&$S_{m}$& 0.5930&&\\
 &$S$&  3.9761&&\\
 primate (5) &$x$&  0.7160& 0.0568& 0.0794\\
  &$B$& 0.5820& 0.0593& 0.1019\\
 &$x_{eff}$& 0.5334& 0.0223& 0.0419\\
 &$S_{m}$& 0.5902& 0.0554& 0.0939\\
  &$S$&  4.0123& 0.0391& 0.0097\\
  \end{tabular}
  \end{table}
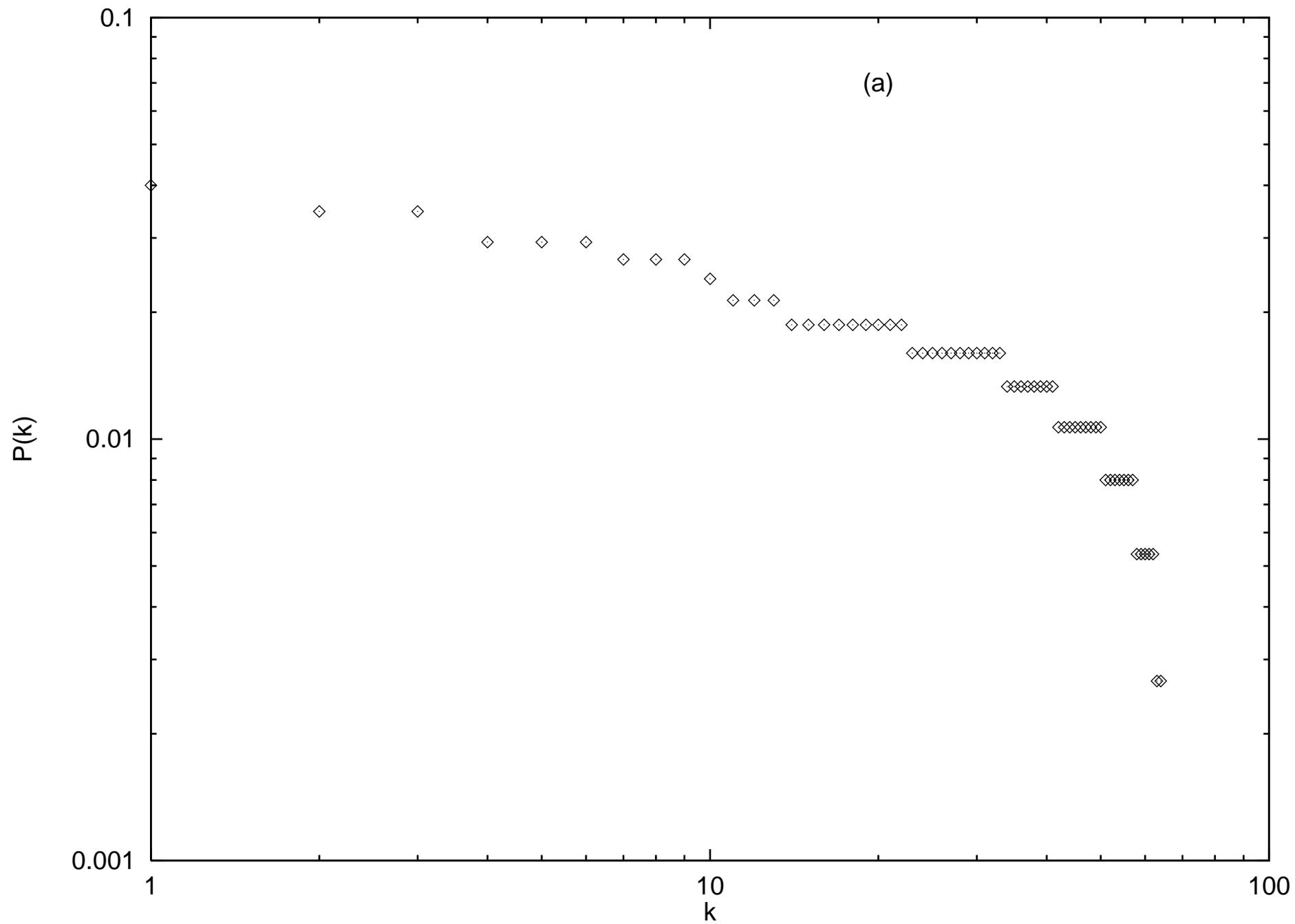
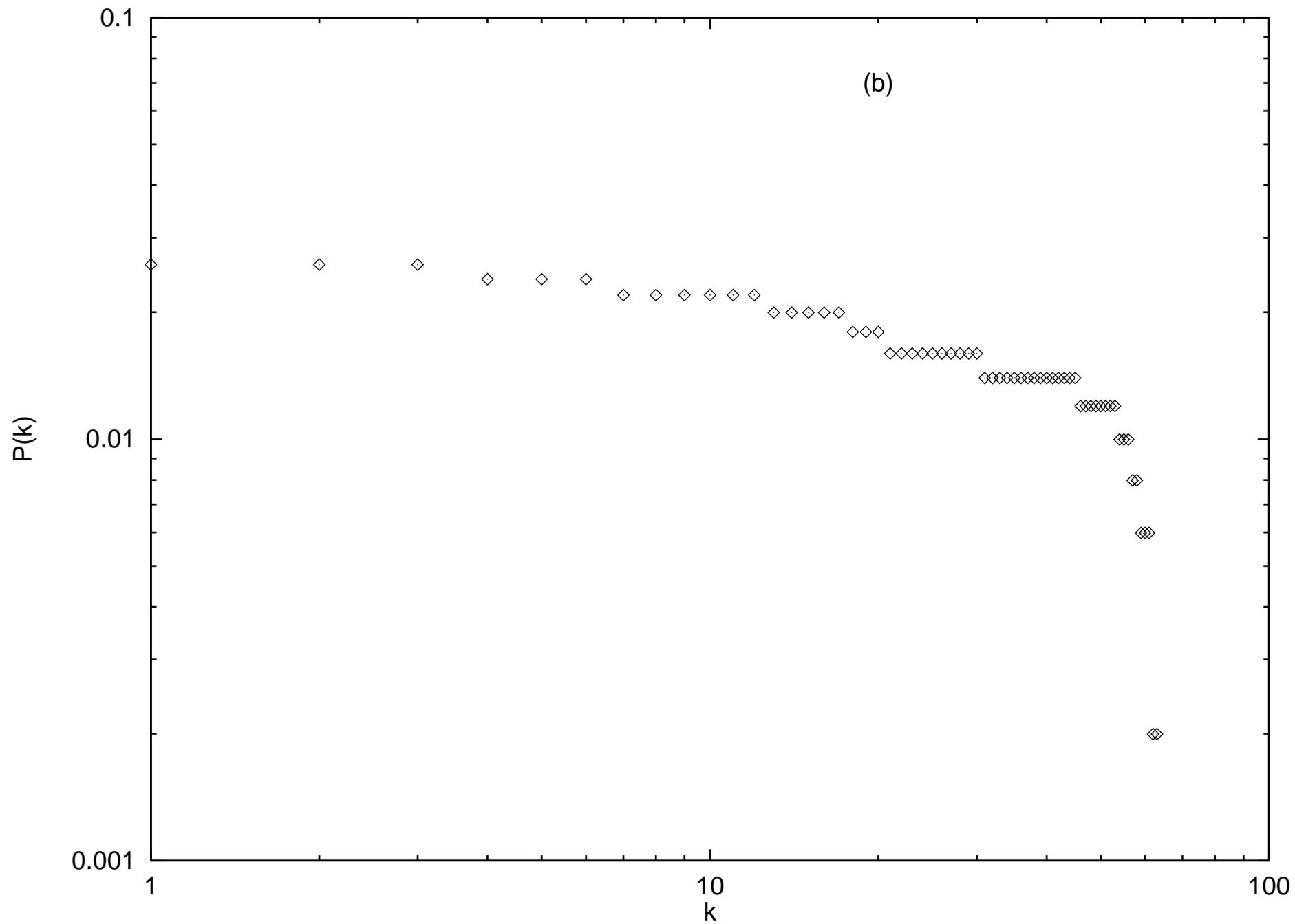
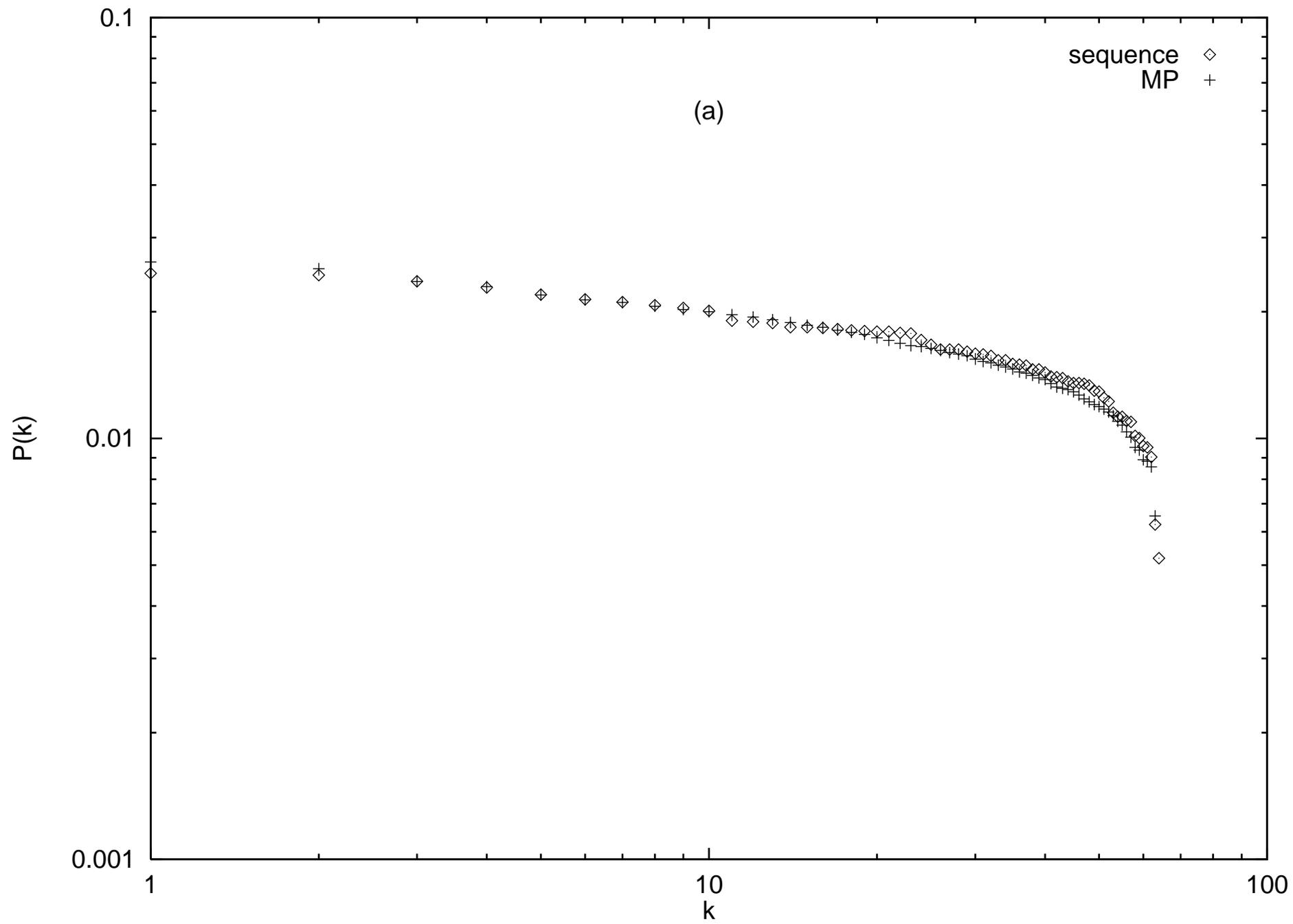
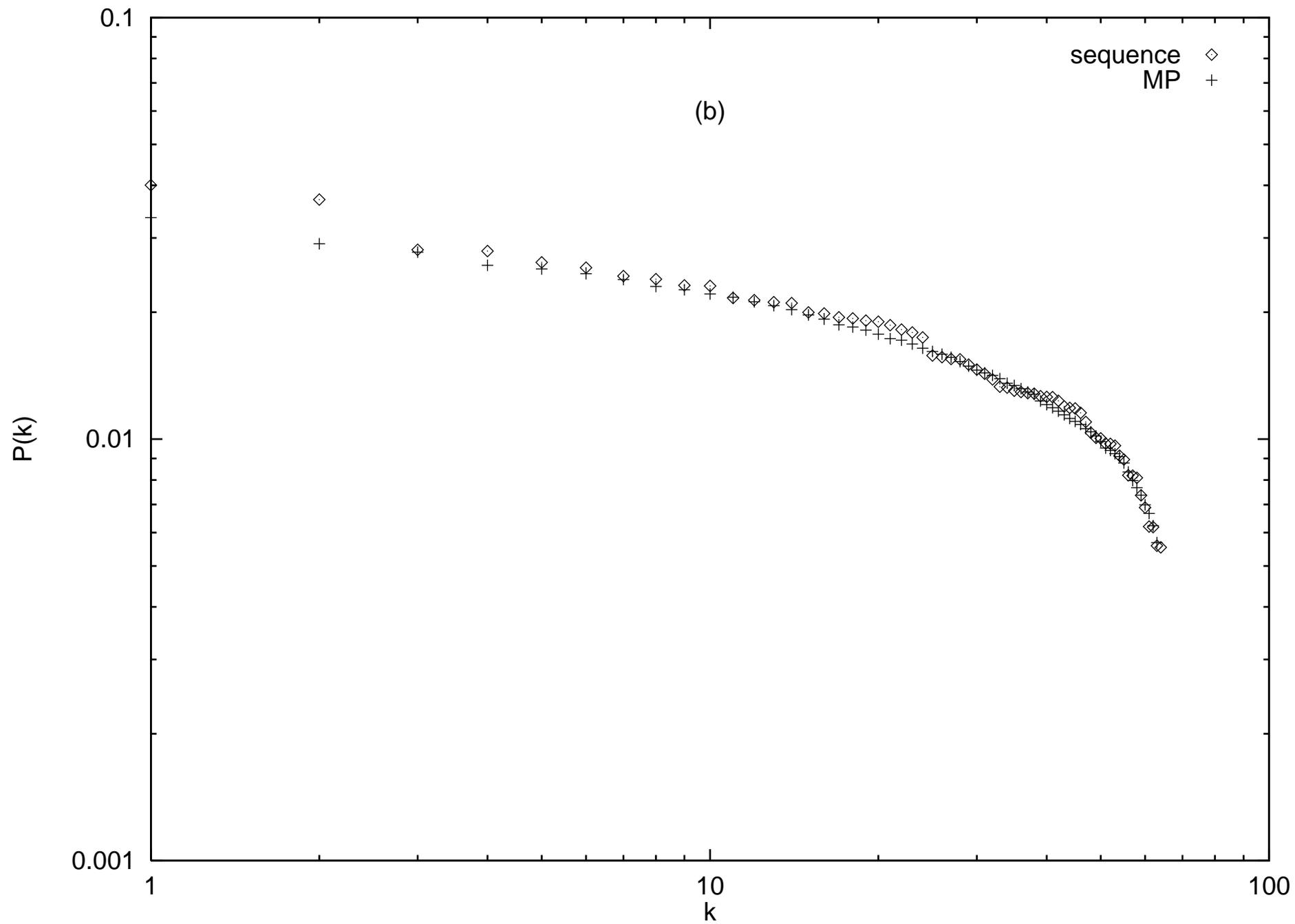
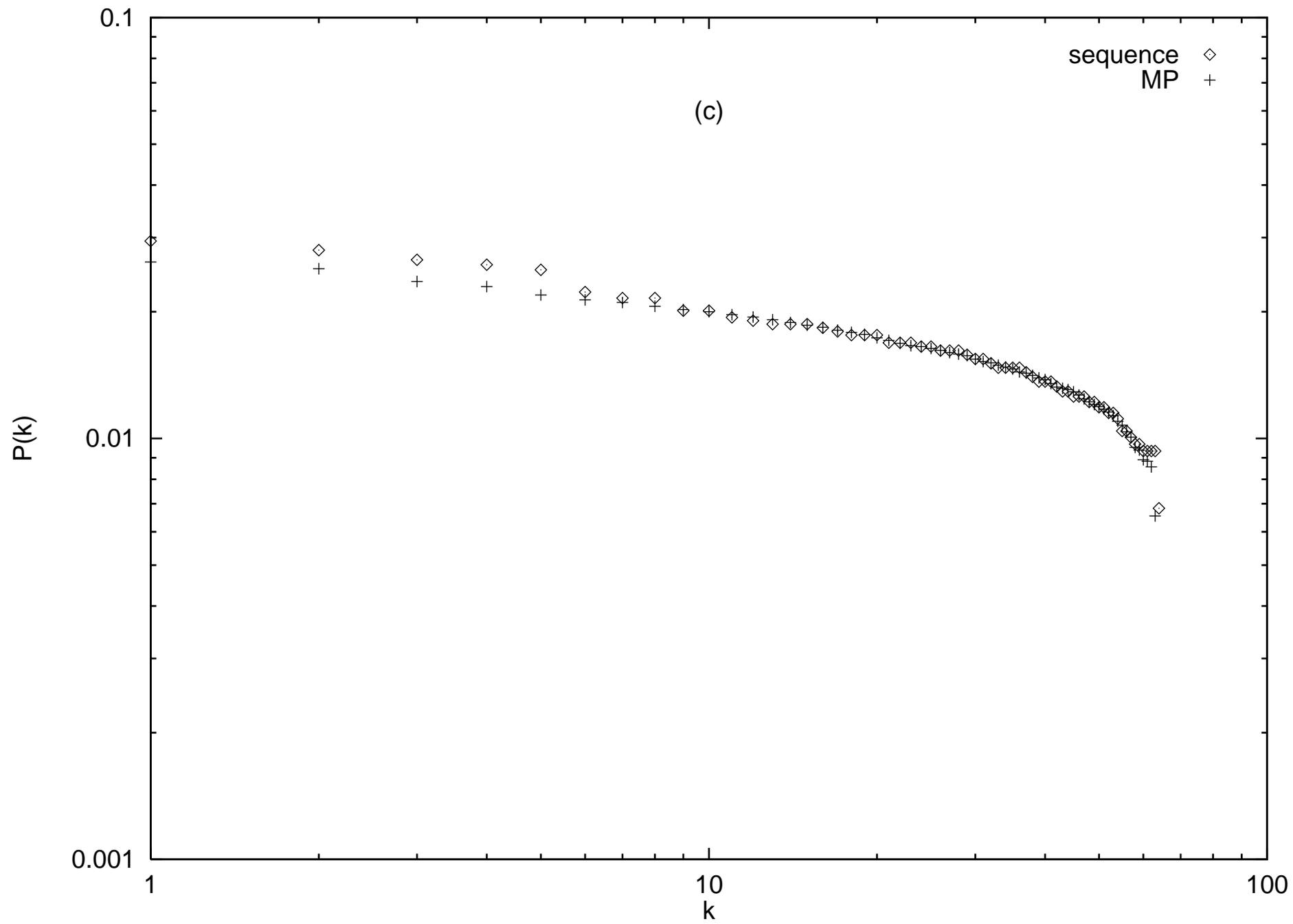
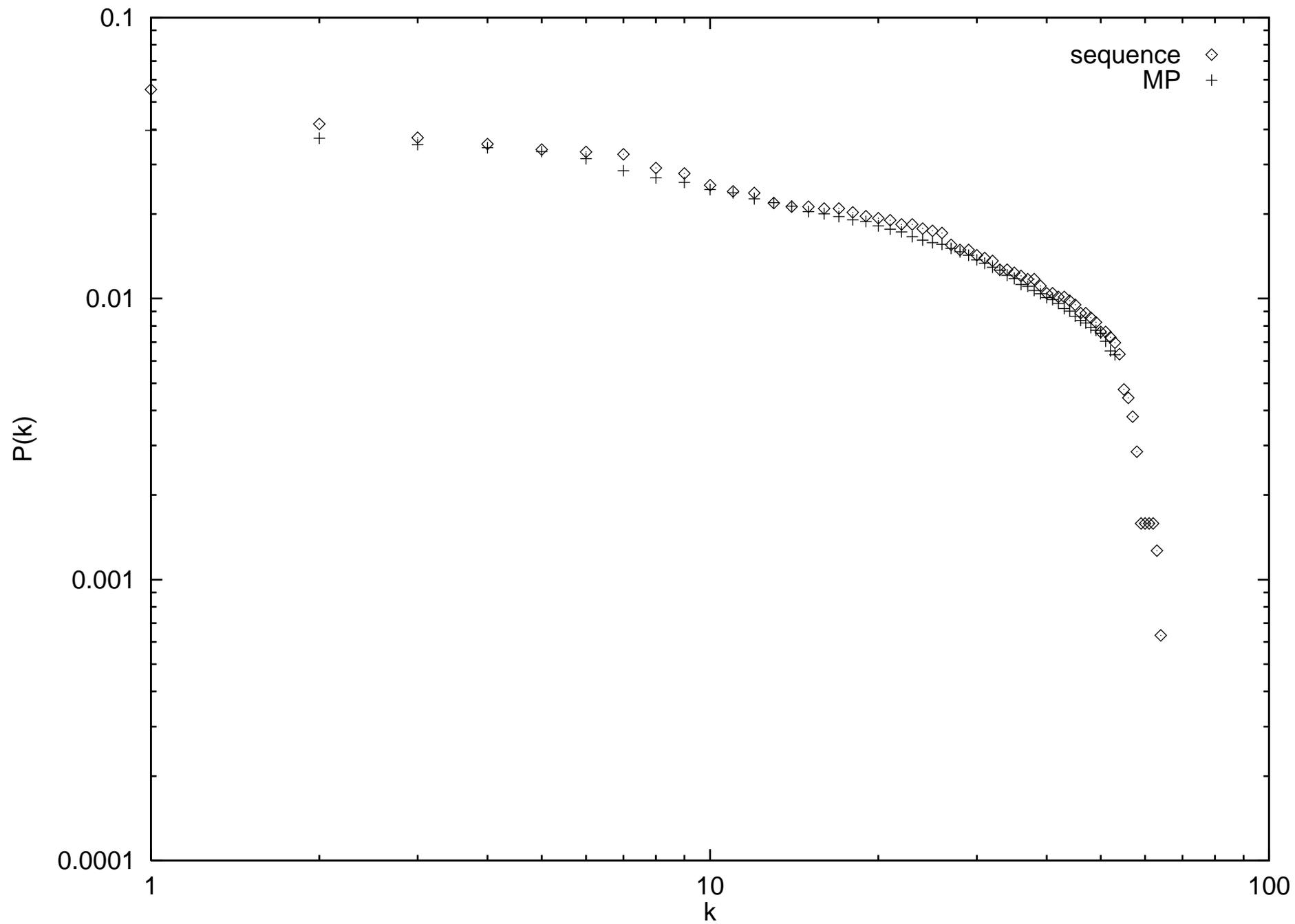
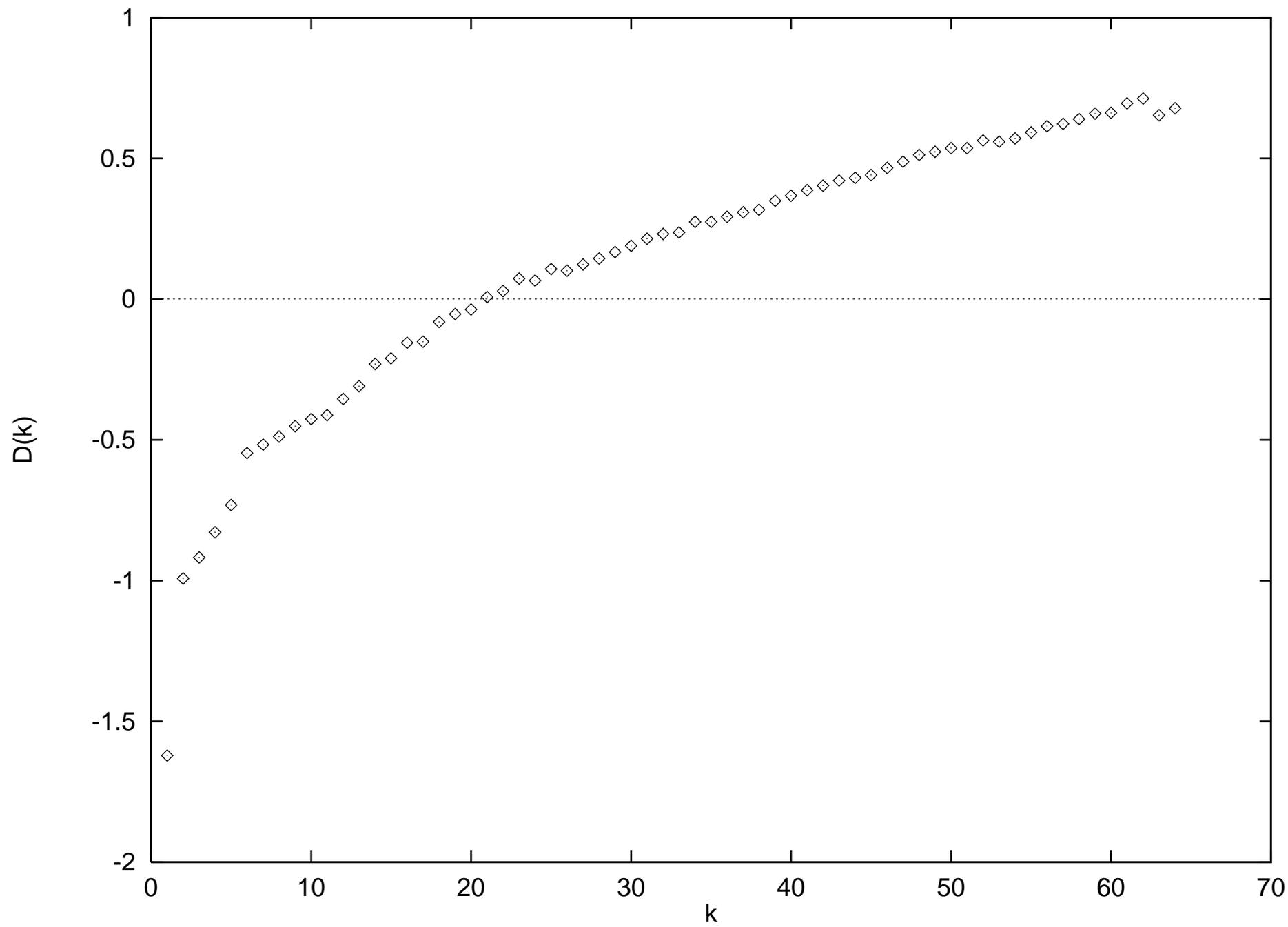
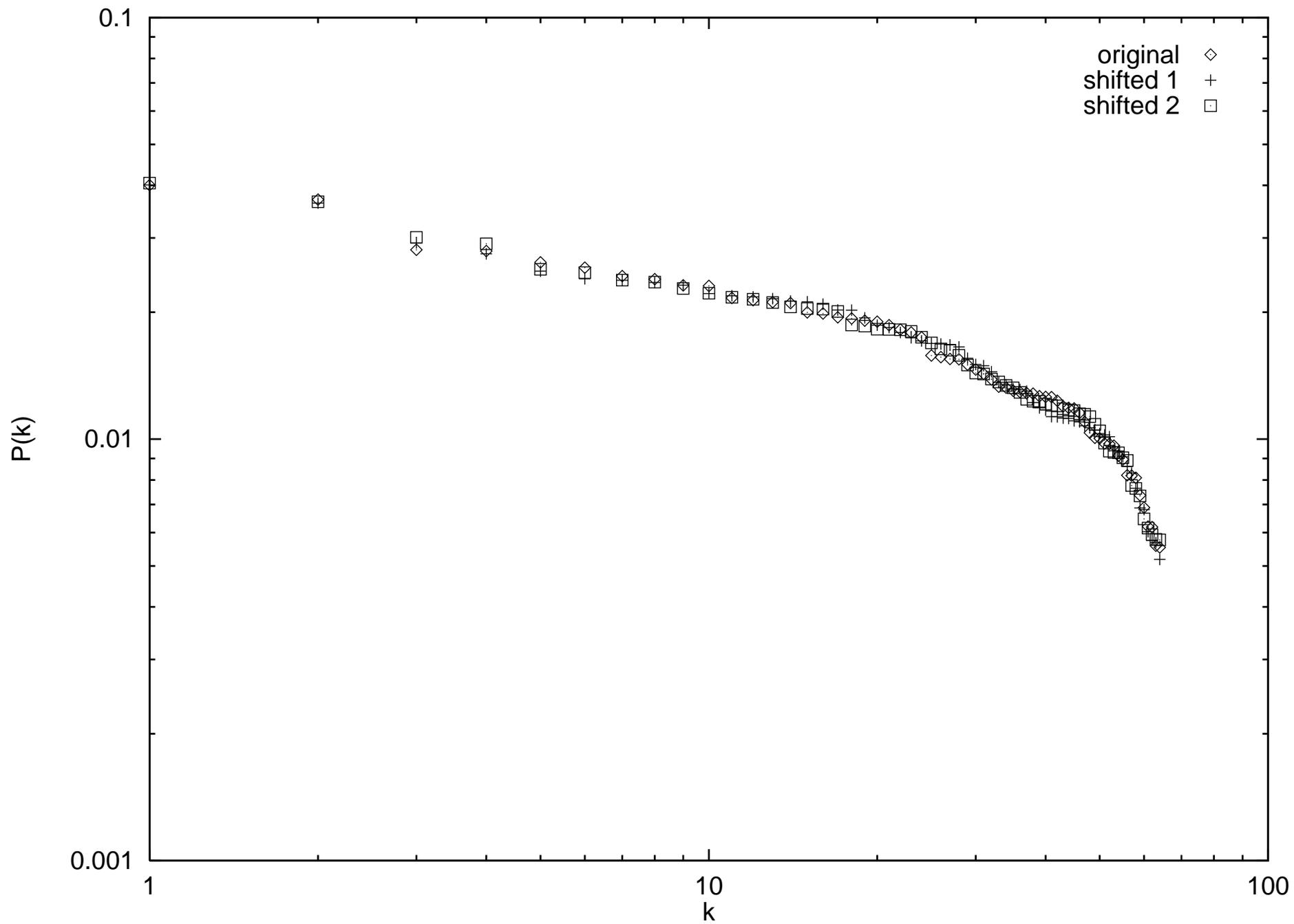
\end{document}